\documentclass{aa}  
\usepackage{graphicx}
\usepackage{txfonts}
\defcitealias{assef22}{A22}
\defcitealias{mathis77}{MRN77}

\begin{document} 

   \title{A Massive Gas Outflow Outside the Line-of-Sight: Imaging Polarimetry of the Blue Excess Hot Dust Obscured Galaxy W0204--0506}

   \author{
        R.~J. Assef\inst{\ref{inst1}} \and
        M. Stalevski\inst{\ref{inst2},\ref{inst3}} \and
        L.~Armus\inst{\ref{inst4}} \and       
        F.~E. Bauer\inst{\ref{inst5},\ref{inst6},\ref{inst7},\ref{inst8}} \and
        A. Blain\inst{\ref{inst9}} \and
        M. Brightman\inst{\ref{inst10}} \and
        T. D\'iaz-Santos\inst{\ref{inst11},\ref{inst12}} \and
        P.~R.~M. Eisenhardt\inst{\ref{inst13}} \and
        R. Fern\'andez-Aranda\inst{\ref{inst11},\ref{inst14}} \and
        H.~D. Jun\inst{\ref{inst15}} \and
        M. Liao\inst{\ref{inst1}} \and
        G. Li\inst{\ref{inst16},\ref{inst18},\ref{inst1}} \and
        L.~R. Martin\inst{\ref{inst8}} \and
        E. Shablovinskaia\inst{\ref{inst1}} \and
        D. Shobhana\inst{\ref{inst1}} \and
        D. Stern\inst{\ref{inst12}} \and
        C.-W. Tsai\inst{\ref{inst19},\ref{inst16},\ref{inst20}} \and
        A. Vayner\inst{\ref{inst4}} \and
        D.~J. Walton\inst{\ref{inst21}} \and
        J. Wu\inst{\ref{inst18},\ref{inst16}} \and
        D. Zewdie\inst{\ref{inst22}}
    }

   \institute{
        Instituto de Estudios Astrof\'isicos, Facultad de Ingenier\'ia y Ciencias, Universidad Diego Portales, Av. Ej\'ercito Libertador 441, Santiago, Chile.~\email{roberto.assef@mail.udp.cl}\label{inst1}
        \and
        Astronomical Observatory, Volgina 7, 11060 Belgrade, Serbia\label{inst2}
        \and
        Sterrenkundig Observatorium, Universiteit Gent, Krijgslaan 281-S9, Gent B-9000, Belgium\label{inst3}
        \and
        Caltech/IPAC, 1200 E. California Blvd. Pasadena, CA 91125, USA\label{inst4}
        \and
        Instituto de Astrof\'isica, Facultad de F\'isica, Pontificia Universidad Cat\'olica de Chile, Av. Vicu\~na Mackenna 4860, Macul Santiago, 7820436, Chile\label{inst5}
        \and
        Centro de Astroingenier\'ia, Facultad de F\'isica, Pontificia Universidad Cat\'olica de Chile, Av. Vicu\~na Mackenna 4860, Macul Santiago, 7820436, Chile\label{inst6}
        \and
        Millennium Institute of Astrophysics, Nuncio Monse\~nor S\'otero Sanz 100, Of 104, Providencia, Santiago, Chile\label{inst7}
        \and
        Space Science Institute, 4750 Walnut Street, Suite 205, Boulder, CO 80301, USA\label{inst8}
        \and
        Department of Physics and Astronomy, University of Leicester, University Road, Leicester LE1 7RH, UK\label{inst9}
        \and
        Cahill Center for Astronomy and Astrophysics, California Institute of Technology, 1200 E. California Boulevard, Pasadena, 91125, CA, USA\label{inst10}
        \and
        Institute of Astrophysics, Foundation for Research and Technology-Hellas (FORTH), Heraklion, 70013, Greece\label{inst11}
        \and
        School of Sciences, European University Cyprus, Diogenes street, Engomi, 1516 Nicosia, Cyprus\label{inst12}
        \and
        Jet Propulsion Laboratory, California Institute of Technology, Pasadena, CA 91109, USA\label{inst13}
        \and
        Department of Physics, University of Crete, 70013, Heraklion, Greece\label{inst14}
        \and
        Department of Physics, Northwestern College, 101 7th St SW, Orange City, IA 51041, USA\label{inst15}
        \and
        National Astronomical Observatories, Chinese Academy of Sciences, 20A Datun Road, Chaoyang District, Beijing 100101, China\label{inst16}
        \and
        University of Chinese Academy of Sciences, Beijing 100049, People's Republic of China\label{inst18}
        \and
        School of Astronomy and Space Science, University of Chinese Academy of Sciences, Beijing 100049, China\label{inst19}
        \and
        Institute for Frontiers in Astronomy and Astrophysics, Beijing Normal University, Beijing 102206, China\label{inst20}
        \and
        Centre for Astrophysics Research, University of Hertfordshire, College Lane, Hatfield AL10 9AB, UK\label{inst21}
        \and
        Centre for Space Research, North-West University, Potchefstroom 2520, South Africa\label{inst22}   
    }

   \date{}

 
  \abstract
   {}
   {Hot Dust Obscured Galaxies (Hot DOGs) are a population of hyper-luminous, heavily obscured quasars. Although nuclear obscurations close to Compton-thick are typical, a fraction show blue UV spectral energy distributions consistent with unobscured quasar activity, albeit two orders of magnitude fainter than expected from their mid-IR luminosity. The origin of the UV emission in these Blue excess Hot DOGs (BHDs) has been linked to scattered light from the central engine. Here we study the properties of the UV emission in the BHD WISE~J020446.13--050640.8 (W0204--0506).}
   {We use imaging polarization observations in the $R_{\rm Special}$ band obtained with the FORS2 instrument at VLT. We compare these data with radiative transfer simulations to constrain the characteristics of the scattering material.}
   {We find a spatially integrated polarization fraction of $24.7\pm 0.7$\%, confirming the scattered-light nature of the UV emission of W0204--0506. The source is spatially resolved in the observations and we find a gradient in polarization fraction and angle that is aligned with the extended morphology of the source found in {\it{HST}}/WFC3 imaging. A dusty, conical polar outflow starting at the AGN sublimation radius with a half-opening angle $\lesssim 50~\rm deg$ viewed at an inclination $\gtrsim 45~\rm deg$ can reproduce the observed polarization fraction if the dust is graphite-rich. We find that the gas mass and outflow velocity are consistent with the range of values found for [O{\sc~iii}] outflows through spectroscopy in other Hot DOGs, though it is unclear whether the outflow is energetic enough to affect the long-term evolution of the host galaxy. Our study highlights the unique potential for polarization imaging to study dusty quasar outflows, providing complementary constraints to those obtained through traditional spectroscopic studies.}
   {}
   \titlerunning{A Massive Gas Outflow Outside the Line-of-Sight}
   \maketitle

%

\section{Introduction}

Quasar activity is thought to be important in regulating, and eventually suppressing, the stellar growth of massive galaxies \citep[e.g.,][]{dimatteo05}. During this period of intense accretion, the quasar is expected to drive massive outflows that will heat up the interstellar medium (ISM), lowering the efficiency of star-formation (SF), and depleting the gas reservoir of the host galaxy, limiting its future SF history. Such effects should be more noticeable in the most luminous quasars known, and indeed many studies have found massive ionized gas outflows in such objects \citep[e.g.,][]{harrison16,zakamska16,bischetti17,temple19,vayner24a}.

Hot Dust Obscured Galaxies \citep[Hot DOGs;][]{eisenhardt12,wu12} are some of the most luminous \citep[$L_{\rm Bol}>10^{13}L_{\odot}$, with 10\% exceeding $L_{\rm Bol}>10^{14}L_{\odot}$;][]{tsai15}, obscured \citep[$2.5\lesssim E(B-V)\lesssim 30$;][]{assef15} quasars known in the Universe, and indeed display an array of features that suggest strong feedback is occurring. They seem to have large supermassive black hole (SMBH) masses for their stellar masses as well as higher Eddington ratios compared to similarly luminous type 1 quasars \citep{li24}, potentially increasing the effects of the quasar on the host galaxy. Hot DOGs show massive, fast (up to $\sim$10,000~$\rm km~\rm s^{-1}$) outflows detected in the [O{\sc iii}]$\lambda5007$\AA\ emission line \citep{jun20,finnerty20,vayner24b} with mass outflow rates up to $\sim$10$^{4}~M_{\odot}~\rm yr^{-1}$ as well as slower yet powerful outflows detected in the [C{\sc ii}]$\lambda158\mu$m emission line \citep{diaz-santos16, liao24a}. These outflows may be powering the strong turbulence found in the FIR emission line across the host galaxies of these obscured quasars \citep{diaz-santos16, diaz-santos21, liao24b}.

Analysis of Hot DOG spectral energy distributions (SEDs) using the SED templates of \citet{assef10} shows that some of them have blue emission in excess of the maximum SF allowed by those templates, indicating that in addition to the hyper-luminous, heavily obscured quasar needed to account for the bright mid-IR emission, a second unobscured or lightly obscured template with $\sim1\%$ the luminosity of the obscured component is required \citep{assef16}. We refer to these objects as Blue excess Hot DOGs (BHDs), and \citet{li24} have recently found that BHDs may account for as much as 25\% of all Hot DOGs. Detailed studies of a few objects \citep{assef16, assef20} concluded that this blue emission originates from scattered light from the hyperluminous obscured quasar. Imaging polarimetry of one of these objects (WISE J011601.41--050504.0, W0116--0505 hereafter) found its UV continuum to have a polarization fraction of 10.8$\pm$1.9\% \citep[][\citetalias{assef22} hereafter]{assef22}, consistent with the scattering scenario. Similar conclusions have been reached for other objects displaying similar SEDs \citep{alexandroff18, stepney24}. 

In this article we present imaging polarimetric observations for another BHD, WISE J020446.13--050640.8 (W0204--0506 hereafter). In contrast to the results for W0116--0505 by \citetalias{assef22}, our observations for W0204--0506 spatially resolve the source, providing a map of the structures responsible for the scattering. In \S\ref{sec:obs} we present the photometric, spectroscopic and polarimetric observations we use in this paper. In \S\ref{sec:pol_measurements} we present the linear polarization fraction and angle of the source, both spatially integrated and resolved, and in \S\ref{sec:discussion} we interpret these observations. In a companion paper \citep{assef24b}, we present spatially unresolved polarimetric observations for a sample of BHDs. We assume a concordance flat $\Lambda$CDM cosmology with $H_0=70~\rm km~\rm s^{-1}~\rm Mpc$ and $\Omega_{\rm M}=0.3$ throughout the article.

\section{Observations}\label{sec:obs}

\subsection{Imaging Polarimetry}

Imaging polarimetry observations of W0204--0506 in the $R_{\rm Special}$ band were obtained using the FOcal Reducer/low dispersion Spectrograph 2 (FORS2) instrument at the Very Large Telescope (VLT) as part of the program {\tt{111.24UL}}. The observations were divided into two observing blocks (OBs), carried out on the nights of UT 2023 September 20 and 22 under clear conditions with average seeing values of 0.6\arcsec\ and 0.8\arcsec, respectively. Observations on both nights where obtained with a mean airmass of 1.1. Each observing block consisted of 2$\times$353~s exposures at retarder plate angles of 0, 22.5, 45, and 67.5~deg. Observations of standard stars using the same set of retarder plate angles are regularly obtained by the observatory. We also analyze the observations of the zero polarization standards WD 2039--202 and WD 2359--434 obtained on the night of UT 2023 September 14, as well as observations of the polarization standard Vela1 95 obtained on the night of UT 2023 October 21. The latter is the polarization standard observed closest in time to W0204--0506 that was not saturated or heavily out of focus. 

We follow \citetalias{assef22} and used the {\tt{EsoRex}}\footnote{\url{https://www.eso.org/sci/software/cpl/esorex.html}} pipeline to subtract the bias. Cosmic rays were removed using the Python package Astro-SCRAPPY \citep{astroscrappy} which is based on the algorithm of \citet{vandokkum01}.

The polarization fraction and angle are generally obtained following the approach outlined by \citetalias{assef22}, which follows that outlined in the FORS2 User Manual and includes the background polarization corrections from \citet{gg20}. In short, we use the flux of the source measured in the ordinary and extraordinary beams, $f^o(\theta_i)$ and $f^e(\theta_i)$ for a retarder plate angle of $\theta_i$, to estimate the Stokes $Q$ and $U$ parameters. The polarization fraction is then given by

\begin{equation}\label{eq:P}
p = \sqrt{Q^2+U^2} ,
\end{equation}

\noindent and the polarization angle by 

\begin{equation}\label{eq:chi}
\chi = \frac{1}{2} \arctan{\left(\frac{U}{Q}\right)} .
\end{equation}

An additional step we have taken here is to align and PSF-match the images before extracting $f^o(\theta_i)$ and $f^e(\theta_i)$ to ensure our results are not affected by pointing jitter or changes in the PSF. For the target source these corrections are small, as the pointing jitter is less than a pixel and the PSF varies by less than 0.1\arcsec\ within an OB. For the science target images we use the {\tt{ImageMatch}}\footnote{https://github.com/obscode/imagematch} software from Carnegie Observatories. This approach does not work properly for the standards as those observations are much shallower, limiting the number of well detected stars in the FoV to estimate the PSF shape. Instead we just convolve the standard star images to a common PSF assuming a 2D Gaussian shape using the {\tt{make\_2dgaussian\_kernel}} function of {\tt{photutils}} and the {\tt{convolve}} routine from {\tt{astropy}}. Note, however, that if we do not apply these PSF-matching steps our results are not qualitatively affected.

\subsection{Spectroscopy and Broad-band Photometry}\label{ssec:phot_spec}

Spectroscopic observations of W0204--0506 obtained with the GMOS-S instrument at the Gemini South Observatory are presented in \citet{assef16}. The source shows broad C{\sc iv}$\lambda$1549\AA\ and C{\sc iii}]$\lambda$1909\AA\ emission lines, which are typical of type 1 quasars. However, the spectrum also displays bright He{\sc ii}$\lambda$1640\AA\ emission with an equivalent width comparable to that of C{\sc iv} and C{\sc iii}], which is unusual in type 1 quasars \citep[e.g.,][]{vandenberk01}, but common in radio galaxies \citep[e.g.,][]{stern99}. With an updated procedure to improve the error spectrum that will be presented in an upcoming study by \citet{eisenhardt24}, we have modeled the emission lines using single Gaussian profiles with a power-law continuum, similar to the approach described in \citet{assef20}. Given that broad UV emission lines are usually seen to be blue-shifted in type 1 quasars in general \citep[see, e.g.,][]{decarli18} and in Hot DOGs in particular \citep{diaz-santos21}, we use the narrower He{\sc ii} emission line alone to determine the redshift for this source, $z=2.0993\pm 0.0003$. For reference, with respect to this redshift, the C{\sc iv} and C{\sc iii}] emission lines are blue-shifted by about $200~\rm km~\rm s^{-1}$ and $50~\rm km~\rm s^{-1}$ respectively. We measure FWHM velocities of $1616^{+212}_{-173}~\rm km~\rm s^{-1}$ for C{\sc iv}, $616^{+73}_{-48}~\rm km~\rm s^{-1}$ for He{\sc ii}, and $870^{+417}_{-285}~\rm km~\rm s^{-1}$ for C{\sc iii}], which are broadly consistent with those reported by \citet{assef16}.

\begin{table}
    \caption{\label{tab:phot} Photometric Data}
    \begin{small}
    \begin{tabular}{lccl}
        \hline \hline\\
        Band & Flux & Uncertainty & Source\\
             & ($\mu$Jy) & ($\mu$Jy) & \\
        \hline\\
        $u^{\prime}$  &  \phantom{0}\phantom{0}\phantom{0}\phantom{0}2.2 & \phantom{0}\phantom{0}1.3 & SDSS DR17\\
        $g^{\prime}$  &  \phantom{0}\phantom{0}\phantom{0}\phantom{0}3.1 & \phantom{0}\phantom{0}0.5 & SDSS DR17\\
        $r^{\prime}$  &  \phantom{0}\phantom{0}\phantom{0}\phantom{0}3.6 & \phantom{0}\phantom{0}0.8 & SDSS DR17\\
        $r$           &  \phantom{0}\phantom{0}\phantom{0}\phantom{0}4.1 & \phantom{0}\phantom{0}0.6 & \citet{assef20}\\
        $i^{\prime}$  &  \phantom{0}\phantom{0}\phantom{0}\phantom{0}6.9 & \phantom{0}\phantom{0}1.1 & SDSS DR17\\
        $z^{\prime}$  &  \phantom{0}\phantom{0}\phantom{0}\phantom{0}4.2 & \phantom{0}\phantom{0}4.2 & SDSS DR17\\
        $J$           &  \phantom{0}\phantom{0}\phantom{0}\phantom{0}8.0 & \phantom{0}\phantom{0}1.6 & \citet{assef15}\\
        $H$           &            \phantom{0}\phantom{0}\phantom{0}32.8 & \phantom{0}\phantom{0}7.4 & VHS DR5\\
        $Ks$          &            \phantom{0}\phantom{0}\phantom{0}27.8 & \phantom{0}\phantom{0}7.3 & VHS DR5\\
        W1            &            \phantom{0}\phantom{0}\phantom{0}28.9 & \phantom{0}\phantom{0}2.4 & CatWISE2020\\
        $[3.6]$       &            \phantom{0}\phantom{0}\phantom{0}37.2 & \phantom{0}\phantom{0}1.9 & \citet{griffith12}\\
        $[4.5]$       &            \phantom{0}\phantom{0}\phantom{0}52.3 & \phantom{0}\phantom{0}2.6 & \citet{griffith12}\\
        W2            &            \phantom{0}\phantom{0}\phantom{0}53.7 & \phantom{0}\phantom{0}3.9 & CatWISE2020\\
        W3            &                                \phantom{0}2555.5 &                     127.8 & WISE AllSky\\
        W4            &                                          11092.8 &                     715.2 & WISE AllSky\\
        \hline
    \end{tabular}
    \end{small}
\end{table}

\begin{figure}
    \centering
    \includegraphics[width=0.48\textwidth]{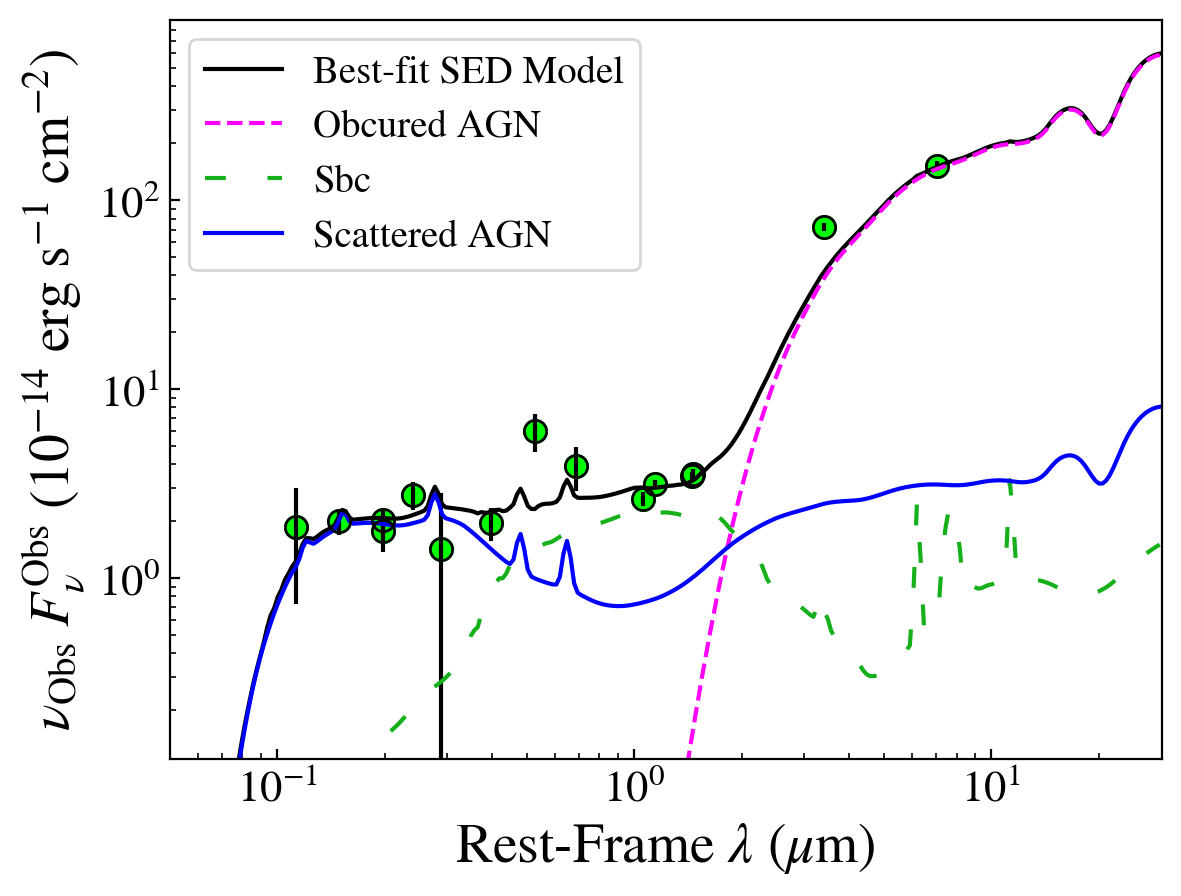}
    \caption{SED of W0204--0506. The green circles show the measured flux densities and their 1$\sigma$ uncertainties in the bands described in the text. The solid black line shows the best-fit SED model composed of a heavily obscured luminous AGN (dashed magenta line), a lightly obscured lower luminosity AGN that accounts for the scattered light component (solid blue line) and a host galaxy from the Sbc template (dashed green line).}
    \label{fg:SED}
\end{figure}

Figure \ref{fg:SED} shows the broad-band SED of the source constructed from combining the profile-fitting fluxes of the WISE W1 and W2 bands from the CatWISE2020 catalog \citep{marocco21} and the W3 and W4 bands from the WISE AllSky Data Release \citep{cutri12} with the {\tt{modelmag}} fluxes of the $u^{\prime}$$g^{\prime}$$r^{\prime}$$i^{\prime}$$z^{\prime}$ bands from the Sloan Digital Sky Survey (SDSS) DR17 \citep{sdssdr17}, the Petrosian aperture fluxes of the $H$ and $Ks$ bands from the Vista Hemisphere Survey (VHS) DR5 \citep{vhsdr5} and the fixed aperture fluxes of the {\it{Spitzer}}/IRAC [3.6] and [4.5] channels from \citet{griffith12}. Additionally we include photometry from the follow-up $r$-band and $J$-band observations discussed in \citet{assef20} and \citet{assef15} respectively. The fluxes for each band are shown in Table \ref{tab:phot}. Figure \ref{fg:SED} also shows the best-fit SED model obtained using the templates of \citet{assef10}, as described in \citet{assef16}, consisting of one quasar component that is heavily obscured ($E(B-V)=10.0^{+0.2}_{-2.0} ~\rm mag$) and very luminous ($\log{L_{6\mu\rm m}/\rm erg~\rm s^{-1}} = 46.89^{+0.04}_{-0.11}$) that dominates the mid-IR emission, one quasar component that is lightly obscured ($E(B-V)=0.10^{+0.01}_{-0.08} ~\rm mag$) and much less luminous ($\log{L_{6\mu\rm m}/\rm erg~\rm s^{-1}} = 45.00^{+0.05}_{-0.49}$, i.e., about 1.3\% of the luminosity of the obscured component) that dominates the UV/optical emission and an intermediate star-forming host galaxy component that dominates the NIR. The numbers differ slightly from those presented by \citet{assef16} and \citet{assef20} due to the specific photometry used as well as the updated redshift value. The $H$-band photometry from VHS significantly deviates from the best-fit SED model, suggesting that the [O{\sc iii}]$\lambda$5007+4959 emission line has a larger equivalent width than the templates used. Based on the flux density difference, we estimate a combined equivalent width of the [O{\sc iii}] lines of about 1300\AA. This is consistent with the large outflows detected in Hot DOGs through this emission line by \citet{jun20} and \citet{finnerty20}. We note that this has only a minor effect in the SED model. If we fit the SED without the $H$-band photometry, the luminosities of each of the AGN components change by less than 1\% and we do not see any changes in the best-fit $E(B-V)$ values. In the next sections we further discuss the presence of outflows in W0204--0506.

\section{Linear Polarization Measurements}\label{sec:pol_measurements}

We first measure the spatially integrated linear polarization fraction and angle for the science target and the standard stars. For this, we measure $f^o(\theta_i)$ and $f^e(\theta_i)$ in each image of each target through a circular aperture using {\tt{photutils}}. We use a 2.0\arcsec\ diameter for W0204--0506, and a  3.0\arcsec\ diameter for the standard stars given the typically worse conditions they were observed in. For Vela1 95 we measure $p=7.79\pm 0.06$\% and $\chi=171.8\pm 0.2~\rm deg$, which are close to the fiducial values of $p=7.89\pm 0.04$\% and $\chi=172.1\pm 0.2$ listed in the FORS2 documentation\footnote{\url{https://www.eso.org/sci/facilities/paranal/instruments/fors/inst/pola.html}}. For the zero polarization standards WD 2039--202 and WD 2359--434 we measure respective polarization fractions of $0.04\pm 0.03$\% and $0.14\pm 0.10$\%, which are formally consistent with no polarization. The differences with the expected values suggest that the systematic errors in our measurements are, at most, a fraction of a percent. 

For W0204--0506 we measure a spatially integrated polarization fraction of $p=24.7\pm 0.7$\% and a polarization angle of $\chi=12.7\pm 0.8~\rm deg$. Splitting the observations by OB, we find consistent values of $p=25.2\pm 1.0$\% and $\chi=12.4\pm 1.1$ for the first night and $p=24.3\pm 0.8$\% and $\chi=13.0\pm 1.0$ for the second night. This is a much higher polarization fraction than the $10.8\pm 1.9$\% measured for W0116--0505 by \citetalias{assef22} and compared to the other BHDs presented in the companion paper \citep{assef24b}, which show a range between $\sim$6\% and $\sim$15\%. Polarization fractions of up to $\sim$15\% have been also found for Extremely Red Quasars (ERQs) by \citet{alexandroff18}. ERQs are closely related to BHDs, as discussed in \citet{assef20}.

\begin{figure*}
    \centering
    \includegraphics[width=\textwidth]{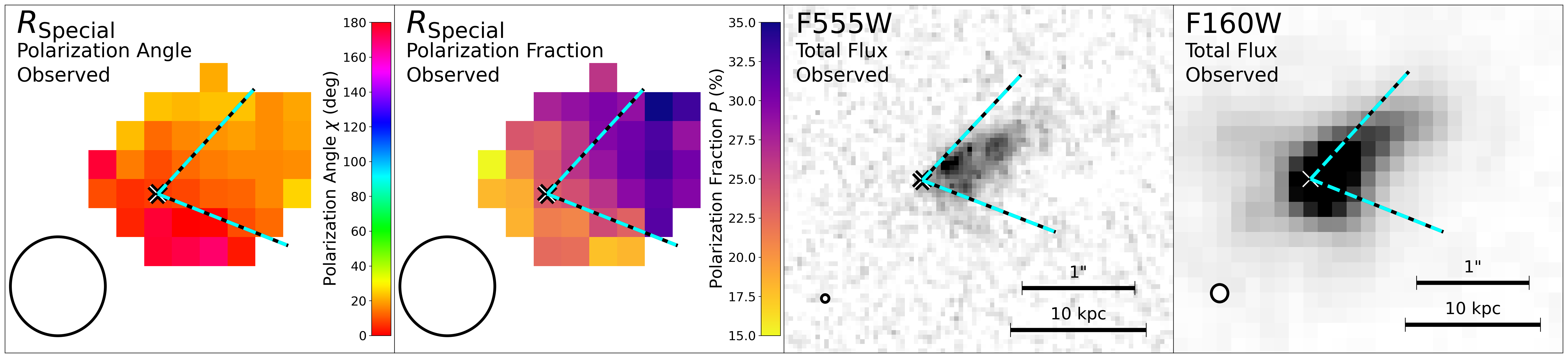}
    \includegraphics[width=\textwidth]{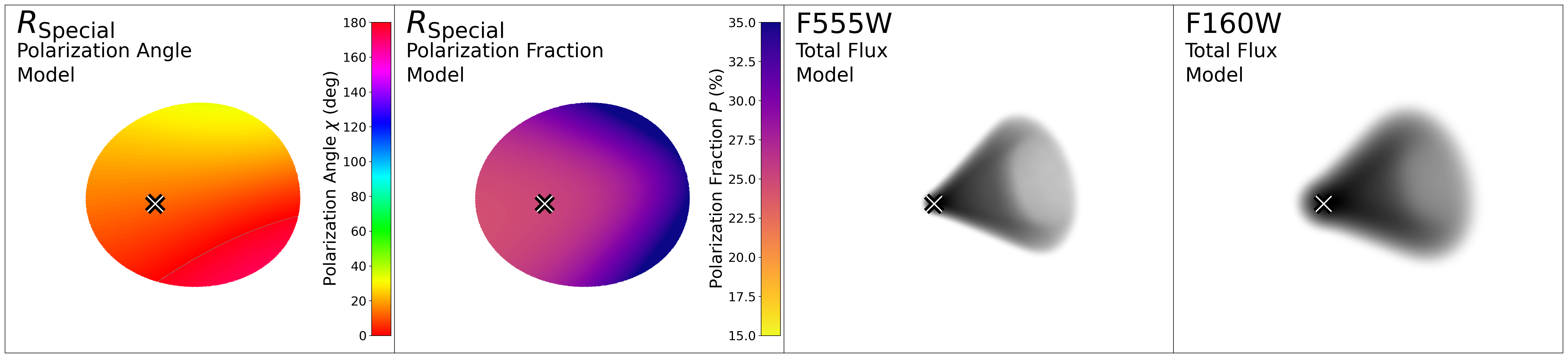}
    \caption{{\it{(Top)}} The two leftmost panels show the resolved maps of the polarization angle and fraction of W0204-0506 measured with FORS2. Polarization values are shown for every pixel where the combined flux from all PSF-matched o- and e-beam images is detected above 5$\sigma$. The two rightmost panels show the {\it{HST}}/WFC3 images in the F555W and F160W bands. All cutouts have a size of 3\arcsec$\times$3\arcsec\ and have been astrometrically aligned using stars within the field of view. The cross shows the position of the brightest F160W pixel in all panels. {\it{(Bottom)}} Representative best-fit model intensity and polarization maps, convolved with the FORS2, F555W or F160W PSF as appropriate. The cross shows the position of the quasar. The cyan-black dashed lines in the top panels show where the edges of a cone with $\psi_{\rm Cone}=40~\rm deg$ and an axis length of 10kpc projected by the 60 degree inclination of the model (see \S\ref{sec:discussion} for details).} 
    \label{fg:W0204_pol_HST}
\end{figure*}

Our observations of W0204--0506 spatially resolve the polarization structure, as shown in Figure \ref{fg:W0204_pol_HST}. A significant gradient is observed in the polarization fraction $p$ from about 20\% in the southeast to about 30\% in the northwest. The typical polarization fraction uncertainty in each pixel is about 4\%. The gradient is also observed in the polarization angle $\chi$, which shifts from about 0 to 20 degrees from the southeast to the northwest, with a typical uncertainty in each pixel of about 3~deg. This direction is aligned with the morphology observed in the {\it{HST}}/WFC3 F555W and F160W imaging presented by \citet{assef20}, as shown in Figure \ref{fg:W0204_pol_HST}. The Figure marks the position of the brightest pixel in the F160W imaging (rest-frame $\sim$5200\AA) in all the bands, likely where the nucleus of the galaxy resides, and it is clear that the bulk of the emission in the F555W band (rest-frame $\sim$1800\AA) is towards the northwest of that position, overlapping with the highest polarization regions in the FORS2 $R_{\rm Special}$ observations (rest-frame $\sim$2100\AA). We note that the contribution from star-formation to the UV emission must be negligible, as it would greatly reduce the polarization fraction; the measured values are too large to suspect dilution as they already exceed what is seen in other highly polarized quasars \citep[e.g.][]{alexandroff18}. Considering the morphology and polarimetry maps, and the prevalence of massive outflows in Hot DOGs in general \citep{jun20, finnerty20}, it seems that the rest-frame UV emission corresponds to light scattered from by dust entrained in a massive outflow almost perpendicular to the line of sight. The gradient in polarization could imply some dilution by starlight around the peak of the F160W flux, but in that case we would not expect a change in the polarization angle. The correlated change in polarization fraction and angle suggest that it could be scattered light from the opposite side of a biconical outflow, possibly obscured or modified by interaction with the ISM. \citetalias{assef22} investigated the possibility that an outflow was acting as the scatterer in the case of W0116--0505 and concluded it was a viable and likely scenario. In the next section we investigate this scenario in more detail for W0204--0506. 

\section{Discussion}\label{sec:discussion}

Recently, \citet{zakamska23} showed that the high polarization fractions measured in ERQs can be explained if the scattering medium is outflowing. They propose that the scattering mechanism is a combination of Thomson scattering from free electrons for the continuum and resonant scattering for the emission lines, and that the scattering medium corresponds to the outer regions of a geometrically thick disk of ionized gas on scales of 10~pc. That physical scale is in stark tension with the $\sim$10~kpc scale of the scattering region in W0204--0506, with the caveat that their model is largely scale independent. The main reason \citet{zakamska23} considered a very compact scale close to the accretion disk for the scattering medium is due to the difficulties in generating large polarization fractions with dust particles, which would be dominant on larger scales (at least at UV wavelengths, see \citealt{young95} for a case where it could be important at NIR wavelengths). Indeed, using the polarization fractions and scattering cross-sections of \citet{draine03} for Milky Way, SMC and LMC dust, we find that in the $R_{\rm Special}$ band (rest $\sim$2100~\AA) SMC and Milky Way type dust can yield a maximum polarization of about 10\%, well below the measured $p=24.7\pm 0.7$\% for the spatially integrated emission of W0204--0506. For LMC-type dust this assessment is more complex as those grains can potentially produce up to $\sim$20\% polarization fractions in backscattering but with very low scattering cross-sections \citep[also see][]{zakamska23}, making it difficult to account for the considerable fraction of AGN light scattered into the line of sight (1.3\%, see \S\ref{ssec:phot_spec}). While the polarization fractions and scattering cross-sections of \citet{draine03} are approximate and somewhat inaccurate blueward of 2700\AA\ due to the phase function approximation used, we conclude it is unlikely that the scattering medium corresponds to typical dust in the ISM of a galaxy. 

We can also rule out Thomson scattering from free electrons as the dominant scattering mechanism, as mentioned earlier, given the large physical size of the scattering medium. This is because dust grains should quickly form beyond the sublimation region of the shock-heated outflow gas. Once they form, dust grains will dominate the scattering properties given their much larger scattering cross-section as compared to free electrons \citepalias[see, e.g., discussion in][]{assef22}. The only possibility left to consider is that the scattering is done by dust with very different properties than those of the \citet{draine03} mixtures. 

To better assess which dust mixtures can reproduce our observations, we use a series of simulations using the code {\tt{SKIRT}}\footnote{\url{https://skirt.ugent.be}} \citep{baes15, camps15, camps20}. {\tt{SKIRT}} is a state-of-the-art code for radiation transfer which employs the Monte Carlo technique to emulate the relevant physical processes such as scattering, absorption and emission, including polarization caused by scattering on spherical dust grains and electrons. Instead of relying on approximations of the phase function, the direction and polarization state of the photons after a scattering event are determined by the Muller matrix elements. Thus, results of the {\tt{SKIRT}} simulations are not subject to the uncertainties mentioned earlier for the \citet{draine03} mixtures. 

We assume a geometry where the scattering medium corresponds to a bi-conical polar outflow wind of half-opening angle $\psi_{\rm Cone}$ where dust is present in the outer 10~\rm deg of the cone, on top of an accretion disk surrounded by an optically thick dust torus of half-opening angle $\psi_{\rm Torus}$. The torus and the polar outflow cones have the same polar axis. The emission of the accretion disk is given by the model of \citet{stalevski16}, and the outflow cone starts at its sublimation radius. The left-most panel of Figure 8 in \citet{stalevski23} shows the general aspects of the geometry we assume here (although we assume a proper cone instead of a hyperboloid). We do not consider a broad line region (BLR) as the $R_{\rm Special}$ broad-band photometry is dominated by the continuum, with only a relatively weak C{\sc iii}] broad emission line contributing in its wavelength range \citep[see][]{assef20}. Note that the dust particles in the outflow cone are not moving with respect to the accretion disk in our model, but that this does not have an impact on our results due to the lack of a BLR. \citet{zakamska23} have shown that this motion could have an impact on the specific polarization properties of the BLR, but not of the accretion disk continuum. We found that the dust optical depth in the explored range does not strongly affect the polarization fraction estimate, but has considerable impact on the fraction of light scattered into the line of sight. For an optical depth of $\tau=0.1$ at rest-frame 5500~\AA\ (i.e., $V$-band), measured for a line of sight from the central engine through the dusty region of the polar outflow cone, an average of order 1\% of the light is scattered, resembling the 1.3\% determined by our SED modeling. We hence fix the optical depth to $\tau=0.1$ for all simulations.    

We use this geometry to test a number of dust mixtures with different grain size distributions and chemical compositions in the cone. We start with the standard Galactic ISM power-law size distribution $\propto a^{-3.5}$ of \citet[][herafter, \citetalias{mathis77}]{mathis77} with a dust grain size range of $a=0.005-0.25~\mu\rm m$. We first consider a combination of silicates and graphites in a 51/49 ratio, as given by the normalization of the grain size distributions in \citet{weingartner01}. We refer to this dust mixture as $\rm MRN77_{\rm gra+\rm sil}$. However, it has recently been suggested that dust in the polar regions of AGN, tentatively associated with dusty winds, is instead dominated by large graphite grains given that it likely originates in the sublimation zone, where small grains and silicate grains are destroyed \citep[see][and references therein]{stalevski17, honig19}. Furthermore, a deficit of silicates is a qualitatively reasonable expectation for the dusty regions of a quasar outflow, as shocks significantly raise the gas temperatures and graphites sublimate at a higher temperature ($\sim$1800~K) than silicates ($\sim$1200~K). Considering this, we also test a graphite-only dust mixture with, otherwise, the same parameters. We refer to this mixture as $\rm MRN77_{\rm gra}$. We also experimented with varying the power-law exponent of the dust distribution in the range of $0$ to $-3.5$ with a step of $-0.5$ and changing the range of grain sizes to larger values of $0.1-1~\mu\rm m$ as well as $1-10~\mu\rm m$. Finally, we tested the \citet{draine03} mixtures for SMC, LMC and Milky Way for completeness, as well as that proposed by \citet{gaskell04}. The \citet{draine03} mixtures have a very similar slope to that of \citetalias{mathis77}, but with an extended grain size range and a smooth high-end cutoff. The dust distribution of \citet{gaskell04} has a flatter grain size distribution dominated by larger grains as compared to that of \citetalias{mathis77}.

The optical properties of the dust grains and elements of the scattering matrix are calculated based on the dielectric function given by \citet{laor93} and \citet{li01}. For simplicity and optimization of the simulations, the dust is distributed smoothly. We note that \citet{marin15} found that fragmentation of the scattering media does not strongly affect the polarization fractions produced by scattering in the polar regions of type 2 AGN, and that uniform dust distributions provide a good approximation in these cases. For the torus composition, we assume in all cases an optically-thick flared disk geometry with a 51/49 silicate-to-graphite ratio and an equatorial optical depth of 10~mag at 9.7$\mu$m. The specific details of the torus composition do not significantly affect the results of the simulations.

\begin{figure*}
    \centering
    \includegraphics[width=\textwidth]{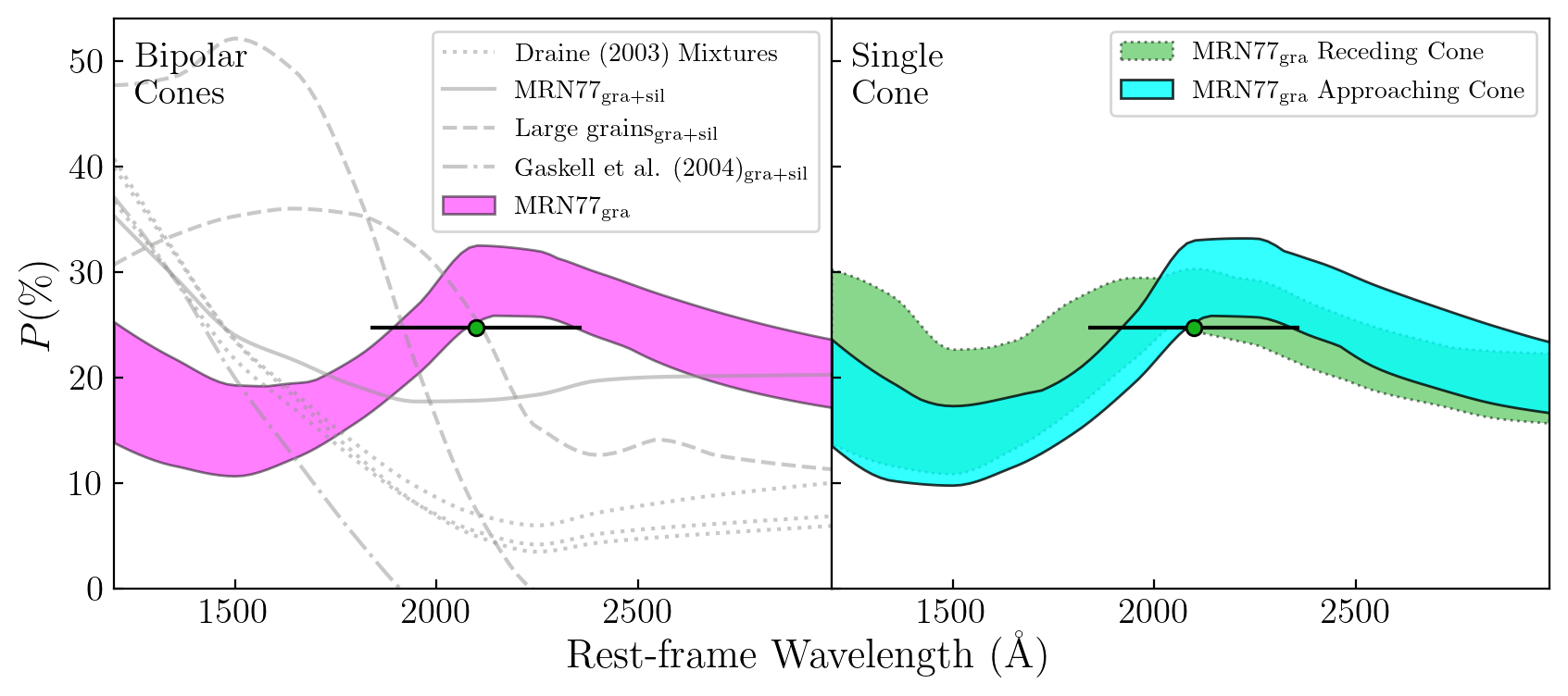}
    \caption{Polarization fraction as a function of wavelength for the different dust models tested with {\tt{SKIRT}} assuming the scattering medium consists of a polar outflow. The left panel shows the expected polarization fraction for bipolar cones for a number of different dust mixtures, while the right panel show the expected values for single approaching or receding cones for the preferred dust mixture. The blue circle shows the measurement for W0204--0506 in the $R_{\rm Special}$ band. The horizontal error-bar shows the width of the photometric band.}
    \label{fg:pfrac_wavelength}
\end{figure*}

All simulations are run for rest-frame wavelengths in the range of 1200$-$3000\AA\ in steps of 150\AA, with additional points close to the effective wavelengths of the $U$$B$$V$$R$$I$ bands for comparison with polarization properties of AGN at low redshifts. Each simulation returns all four Stokes parameters, but the axis of the system is always selected to make the spatially integrated $U=0$. Since we are only dealing with dust scattering there is no circular polarization, and hence $V=0$ by definition. For each simulation, we take the spatially integrated $Q$ and $I$ and convolve them spectrally with the $R_{\rm Special}$ broad-band filter curve (shifted to $z=2.099$). Given the large degeneracies between parameters, we first run simulations with a fixed set of fiducial values of $\psi_{\rm Tor}=50~\rm deg$ and $\psi_{\rm Cone}=30~\rm deg$, varying the inclination angle $\eta$ between 55 and 85~deg in steps of 10~deg to test if we can find polarization fractions that approach the observed value. We find the optimal inclination to match the observations by interpolating within the grid of $\eta$ values. All the dust mixtures that have a graphite-only composition can explain the observed polarization fraction in W0204--0506 regardless of the specific grain size distributions (i.e., both in range and power-law index). Therefore, for the sake of simplicity, when considering graphite-only models in the following, we only consider the standard \citetalias{mathis77} size parameters. On the contrary, nearly all the silicate-graphite mixtures fail, with only the model with the largest grain sizes (i.e., $a=0.1-1~\mu\rm m$ and $a=1-10~\mu\rm m$) being able to reproduce the observed polarization fraction. Figure \ref{fg:pfrac_wavelength} shows the polarization fraction as a function of wavelength for a selection of the silicate/graphite mixtures considered. We focus for the rest of the article on the results for the $\rm MRN77_{\rm gra}$ dust mixture, as it is more physically justified than the model with larger grain sizes. Additionally, we note that in the companion paper that analyzes a larger set of polarization results for BHDs we find that polarization fractions tend to drop towards shorter wavelengths, consistent with $\rm MRN77_{\rm gra}$ dust but not with the mixture with large grain sizes \citep{assef24b}.

\begin{figure*}
    \centering
    \includegraphics[width=\textwidth]{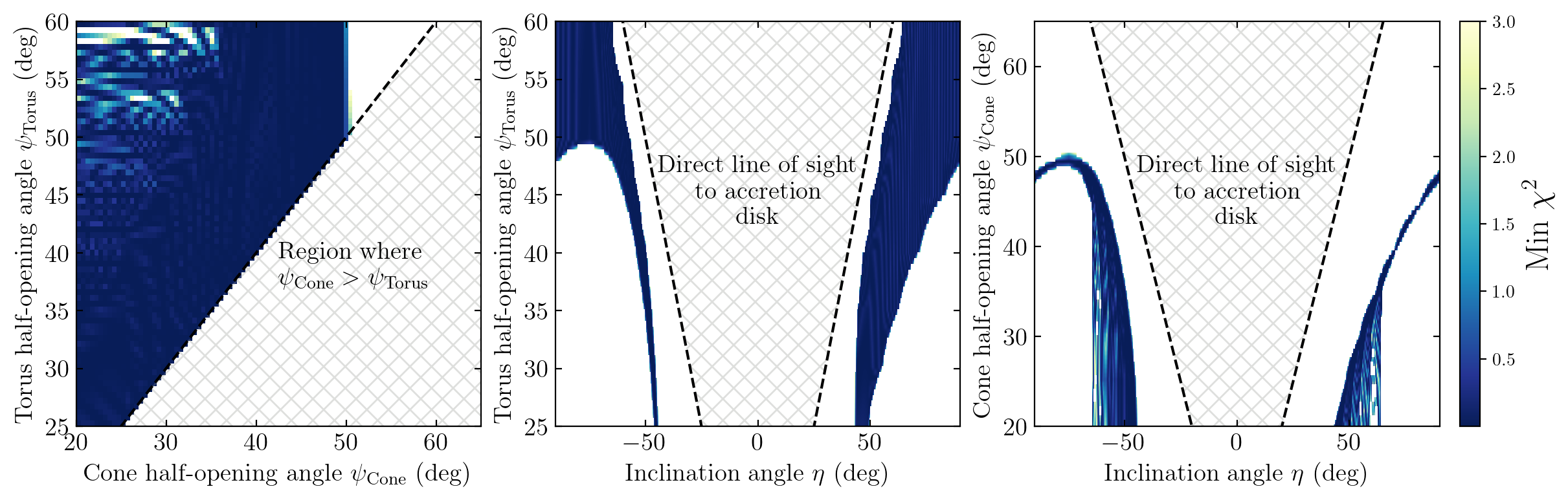}
    \caption{Maps of the minimum $\chi^2$ found as a function of the three parameters fit: $\psi_{\rm Torus}$, $\psi_{\rm Cone}$ and $\eta$. The hatched regions show non-allowed combinations of parameters, either because we would have a direct line of sight to the accretion disk or because the cone would be larger than the torus opening. Approaching (receding) cones are presented as having positive (negative) inclination angles $\eta$.}
    \label{fg:chi2_maps}
\end{figure*}

We then refine the grid for the $\rm MRN77_{\rm gra}$ dust mixture to better map the degeneracies between the parameters. Additionally, given that the images in Figure \ref{fg:W0204_pol_HST} suggest the outflow is only spreading in one direction, we modify the simulations to consist of only one outflow cone that is either approaching or receding from the observer. This makes the simulations somewhat more realistic but note the effects of considering only one cone are fairly small for the the total polarization since the cone approaching the observer is more luminous and hence dominates the estimates (see Figure \ref{fg:pfrac_wavelength}). We compute the simulations in a grid of $\psi_{\rm Torus}=25~\rm deg$ to 60~deg and  $\psi_{\rm Cone}=20~\rm deg$ to $\psi_{\rm Torus}$. In the upper limit of $\psi_{\rm Cone}=\psi_{\rm Torus}$, the cone fills in the aperture of the torus, and only the inner parts of the cone are illuminated by the accretion disk. We also consider inclination angles between $\eta=\psi_{\rm Torus}$ and 90~deg. All parameters are sampled in steps of 5~deg then interpolated. Figure \ref{fg:pfrac_wavelength} shows the range of models that fit the observed polarization with $\chi^2\le 1$. Figure \ref{fg:chi2_maps} shows the degeneracies between the parameters. For display purposes, the approaching cones are presented as positive inclinations while the receding cones are presented as negative inclinations. We constrain the polar outflow cone geometry to have $\psi_{\rm Cone}\lesssim 50~\rm deg$ and the inclination to be $\eta\gtrsim 45~\rm deg$. Furthermore, there is a tight relation observed between the two: the polarization fraction strongly increases with the inclination angle and decreases for broader outflow cones, setting up a trade-off between them. We unfortunately cannot constrain the torus opening angle with the limited data available.

We can also visually compare the properties of a representative model with our observations. We assume an approaching cone with an intermediate inclination of $\eta=60~\rm deg$ for this representative model. We also assume $\psi_{\rm Cone}= 33.5~\rm deg$, estimated from the tight relation with $\eta$ observed in Figure \ref{fg:chi2_maps}. This polar outflow cone shape is shown by the dashed lines in the top panels of Figure \ref{fg:W0204_pol_HST}. We have placed the vertex of the cone at the center of the brightest pixel of the F160W image, which is the likely location of the quasar, and have rotated the cone to match the polarization angle found in the extended UV region. The match with the F555W morphology is very good, particularly at the base, providing additional confirmation that the UV emission in W0204--0506 can be explained by light scattered from a polar outflow. The distribution of polarization angles and fractions that would be observed in the $R_{\rm Special}$ band for this representative model, as well as that of the fluxes observed by {\it{HST}} are shown in the bottom panels of Figure \ref{fg:W0204_pol_HST}, convolved with the appropriate instrument PSFs for comparison and applying the same rotation as before. We assume a torus opening angle of $\psi_{\rm Cone}=40~\rm deg$ for these simulated images, although the results are insensitive to this choice (see Fig. \ref{fg:chi2_maps}). As mentioned earlier, the opposite outflow cone is not immediately observed in the F555W image, which may point to it being dustier or modified, and potentially stalled, by the interaction with the ISM, similar to the scenario recently found for the Hot DOG W2246--0526 by \citet{liao24b} using ALMA observations. As mentioned earlier, a different morphology for this opposite outflow could further account for the gradient observed in the polarization fraction and angle in the FORS2 observations.

Given the cone geometry and the dust optical depth, we can estimate the mass of the outflow by assuming that all the dust involved in the scattering is part of the outflow and that the outflow has a gas composition of pure hydrogen with a uniform distribution throughout the cone, including the dustless inner regions. We also assume the median dust-to-gas ratio of \citet{maiolino01} for AGN of $E(B-V)/N_{\rm H} = 1.5\times 10^{-23}~\rm cm^{2}~\rm mag$ with an $R_V$ of 3.1, and a cone height of 10~kpc. For each model we compute the average scattered flux fraction and linearly scale $\tau_V$ to match the measured 1.3\%. We compute the distribution of the gas mass by adding the values for all models weighted by their likelihood based on the $\chi^2$. We find a 95.4\% range between $5\times 10^{8}~M_{\odot}$ and $4\times 10^{9}~M_{\odot}$. While broad, this gas mass range is consistent with that found for [O{\sc iii}] outflows in Hot DOGs by \citet{finnerty20} of $\sim10^6-10^9~\rm M_{\odot}$, with a median of $\sim2\times 10^7~\rm M_{\odot}$. We further note that a scattered light ray going perpendicularly through the cone at mid-height (i.e., 5~kpc from the base) would suffer an obscuration in the range of $E(B-V)= 0.004~\rm mag$ to $0.04~\rm mag$ (95.4\% confidence), somewhat below the measured value of $E(B-V)=0.1~\rm mag$ from the SED modeling. Hence we would not expect the scattered light to be substantially obscured. We note that in reality the gas and dust are more likely to be concentrated at the base of the cone (which is also strongly suggested by the morphology in the F555W image), which could lead to the larger obscuration measured from the SED modeling.

As discussed in \S\ref{ssec:phot_spec}, the rest-UV spectrum of W0204--0506 shows a bright, relatively narrow He{\sc ii} emission line. This He{\sc ii} line might be produced, at least in part, by the shock-heated gas in the putative outflow. Given the relatively narrow angle of the best-fit cone, if the He{\sc ii} is produced by the outflow we can roughly estimate its intrinsic velocity dispersion by simply correcting for the inclination of the polar outflow cone, $\eta$, which we have constrained to exceed $\sim$45~deg. Doing this we find an intrinsic FWHM $\gtrsim 870~\rm km~\rm s^{-1}$. For the reference inclination of 60~deg assumed earlier, the FWHM would be about $1200~\rm km~\rm s^{-1}$. \citet{finnerty20} found observed FWHM values for [O{\sc iii}] outflows in the range of $1500-8300~\rm km~\rm s^{-1}$, suggesting the inclination of the W0204--0506 outflow cone could be larger than $\sim 65~\rm deg$.

We can also use these gas mass and intrinsic velocity width estimates to assess the energetics of this outflow, namely the mass outflow rate ($\dot{M}_{\rm out}$), the kinetic power of the outflow ($\dot{E}_{\rm out}$) and its momentum flux ($\dot{P}_{\rm out}$). We use here the same equations used by \citet{finnerty20} to study Hot DOG outflows using the [O{\sc iii}] emission line, which depend on the outflow velocity, $v_{\rm out}$, and its gas mass. \citet{finnerty20} estimates $v_{\rm out}$ as $2\sqrt{\sigma^2 + v_{\rm offset}^2}$, where $\sigma$ is the dispersion of the line used to trace the outflow and $v_{\rm offset}$ is its velocity offset. Since we base our redshift measurement for W0204--0506 on the He{\sc ii} emission line (see \S\ref{ssec:phot_spec}) we are not able to estimate $v_{\rm offset}$, so we can only set a lower limit of $v_{\rm out}\gtrsim \sigma_{\rm HeII}/\cos{\eta}$. We find $\dot{M}_{\rm out}\gtrsim 100~M_{\odot}~\rm yr^{-1}$, $\log{\dot{E}_{\rm out}}/{\rm erg~\rm s^{-1}}\gtrsim 43.6$ and $\log{\dot{P}/\rm dyn} \gtrsim 35.9$ with 95.4\% confidence. Taking the fiducial case explored earlier with $\eta=60~\rm deg$, $\psi_{\rm Cone}=33.5~\rm deg$ and $\psi_{\rm Tor}=40~\rm deg$, we instead find larger values for all these quantities, namely $\dot{M}_{\rm out}\sim 430~M_{\odot}~\rm yr^{-1}$, $\log{\dot{E}_{\rm out}}/{\rm erg~\rm s^{-1}}\sim 44.2$ and $\log{\dot{P}/\rm dyn} \sim 36.5$. These are consistent with the range probed by the results of \citet{finnerty20}. \citet{assef16} estimated a total infrared luminosity for W0204--0506 of $L_{\rm TIR}=4.4\times 10^{13}~L_{\odot}$. If we assume this corresponds to the total energy produced by the accretion disk of the AGN, we find that $\dot{E}_{\rm out}/L_{\rm AGN}\sim 0.1\%$ for our fiducial model, although other valid parameter combinations can result in values that even exceed 100\%. The ratio for our fiducial model is somewhat below the typical range of coupling fractions between the AGN luminosity and the outflowing gas that is required by simulations to engage in effective feedback \citep[e.g., see][]{hopkins10, zubovas12}, although drawing firm conclusions from such comparisons is challenging \citep{harrison24}, particularly since much of the AGN-ISM energy coupling might not be easily observable \citep{ward24}.  

The case of W0204--0506 highlights that outflow cones in luminous quasars may be narrow enough, and potentially with enough obscuration at the base, to only be clearly detectable from certain lines of sight when looking for broad forbidden emission lines. We may indeed be missing a larger fraction of the outflow budget in quasars due to inclination effects. The results we have presented here suggest that polarimetric imaging may be a fundamental tool for a comprehensive census of gas outflows in luminous quasars, and to investigate their geometries in a way that is not possible through traditional spectroscopic observations. Further multi-wavelength and high-spatial resolution polarimetric imaging observations of W0204--0506, as well of other Hot DOGs and hyper-luminous quasars, will be critical to provide a complete picture of galaxy evolution and AGN feedback under such extreme conditions.

\begin{acknowledgements}
RJA was supported by FONDECYT grant number 1231718 and by the ANID BASAL project FB210003. DJW acknowledges support from the Science and Technology Facilities Council (STFC; grant code ST/Y001060/1). Part of this research was carried out at the Jet Propulsion Laboratory, California Institute of Technology, under a contract with the National Aeronautics and Space Administration (80NM0018D0004). ML was supported by the grants from the National Natural Science Foundation of China (Nos.11988101, 11973051, 12041302),  the China Postdoctoral Science Foundation (No. 2024M753247) and the International Partnership Program of Chinese Academy of Sciences, Program No.114A11KYSB20210010. TDS acknowledges the research project was supported by the Hellenic Foundation for Research and Innovation (HFRI) under the "2nd Call for HFRI Research Projects to support Faculty Members \& Researchers" (Project Number: 3382).

This research made use of Photutils, an Astropy package for detection and photometry of astronomical sources \citep{photutils}.

\end{acknowledgements}

\bibliographystyle{aa}

\end{document}